\documentclass[a4paper,fleqn,usenatbib,useAMS]{mnras}

\usepackage{graphicx}	% Including figure files
\usepackage{amsmath}	% Advanced maths commands
\usepackage{amssymb}	% Extra maths symbols
\usepackage{multicol}        % Multi-column entries in tables
\usepackage{bm}		% Bold maths symbols, including upright Greek
\usepackage{pdflscape}	% Landscape pages
\usepackage{epsf}
\usepackage{float}
\usepackage{physics}
\usepackage[title]{appendix}
\newcommand{\kms}{\;{\rm km}\,{\rm s}^{-1}}
\newcommand{\kmsmpc}{\kms\;{\rm Mpc}^{-1}}

\newcommand{\HI}{{\sc Hi}}

\newcommand{\hmpc}{h^{-1}{\rm Mpc}}

\newcommand{\msolar}{\;{\rm M}_{\odot}}

\newcommand{\giz}{{\sc Gizmo}}
\newcommand{\muf}{{\sc Mufasa}}

\usepackage[T1]{fontenc}
\usepackage{ae,aecompl}

\usepackage{newtxtext,newtxmath}
\title[Depletion Timescales]{\muf: Timescales for \HI~consumption and SFR depletion of satellite galaxies
in groups}

\author[Rafieferantsoa, Dav\'e \& Naab]{Mika Rafieferantsoa$^{1,2,3}$
\thanks{Contact e-mail: \href{mailto:rafieferantsoamika@gmail.com}
{rafieferantsoamika@gmail.com}}\thanks{South African
Astronomical Observatory, Observatory Road, Cape Town 7925, South Africa},
Romeel Dav\'e$^{4,1,2,5}$, Thorsten Naab$^3$
\\
\\$^1$ University of the Western Cape, Bellville, Cape Town 7535,
South Africa
\\$^2$ South African Astronomical Observatory, Observatory,
Cape Town 7925, South Africa
\\$^3$ Max-Planck-Institut f\"ur  Astrophysik, D-85748 Garching, Germany
\\$^4$ Institute for Astronomy, Royal Observatory, Edinburgh EH9 3HJ, UK
\\$^5$ African Institute for Mathematical Sciences, Muizenberg,
Cape Town 7945, South Africa
}

\date{Last updated 2017 July 6; in original form 2017 July 6}

\pubyear{2017}

\begin{document}
\label{firstpage}
\pagerange{\pageref{firstpage}--\pageref{lastpage}}
\maketitle

% Abstract of the paper
 \begin{abstract}
We investigate the connection between the \HI~content, SFR and
environment of galaxies using a hydrodynamic simulation that incorporates
scaling relations for galactic wind and a heuristic halo mass-based
quenching prescription.  We run two zoom-in simulations of galaxy
groups with $M_{\rm halo}>10^{13}M_\odot$ at $z=0$, selected to
have quiet merger histories.  We track galaxies as they become
satellites, and compute the delay time $\tau_\mathrm{d}$ during
which the satellites are similar to central galaxies at a given
stellar mass, and a fading time $\tau_\mathrm{f}$ during which
satellites go from gas-rich and star-forming to gas-poor and
quiescent.  We find $0.7\lesssim \tau_\mathrm{d}\lesssim 3$~Gyr at
$z=0$, and depends inversely on the satellite halo mass at infall.
At $z\sim1$ we find $\sim0.3\lesssim \tau_\mathrm{d} \lesssim 2$~Gyr,
broadly consistent with a positive correlation with the Hubble time.
For a given halo mass, lower stellar mass galaxies at infall time have higher
$\tau_\mathrm{d}$.  We generally find $\tau_\mathrm{f}\ll
\tau_\mathrm{d}$, ranging between $\sim 150$ Myr at $z\sim0$ and $\sim
80$ Myr at $z\sim1$ based on linear interpolation, with some uncertainty
because they are smaller than our simulation output frequency ($200-300$~Myr).
$\tau_\mathrm{f}$ has no obvious dependency on infall halo mass. Both
timescales show little difference between \HI\ depletion and SF quenching,
indicating that using up the gas reservoir by star formation without refilling
is the main mechanism to transform satellite galaxies at these halo masses.
At a given physical distance from the center of the main halo of interest,
higher redshift galaxies have on average higher cold gas content,
but the ratio of gas (\HI\ or H$_2$) to star formation rate is similar,
indicating that star formation is consistently fed through reservoirs of
\HI\ then H$_2$.  For a given amount of \HI, galaxies have shorter consumption
times in more massive halo structures.  Our results suggest that group-scale
simulations naturally yield a delayed-then-rapid satellite quenching
scenario as inferred from observations both today and at earlier epochs,
though we highlight some quantitative discrepancies.
\end{abstract}

% Select between one and six entries from the list of approved keywords.
% Don't make up new ones.
\begin{keywords}
galaxies: evolution -- galaxies: statistics -- methods: N-body simulations
\end{keywords}

%%%%%%%%%%%%%%%%%%%%%%%%%%%%%%%%%%%%%%%%%%%%%%%%%%

%%%%%%%%%%%%%%%%% BODY OF PAPER %%%%%%%%%%%%%%%%%%

\section{Introduction}

\label{intro}

\begin{figure*}
\includegraphics[width=7.5in]{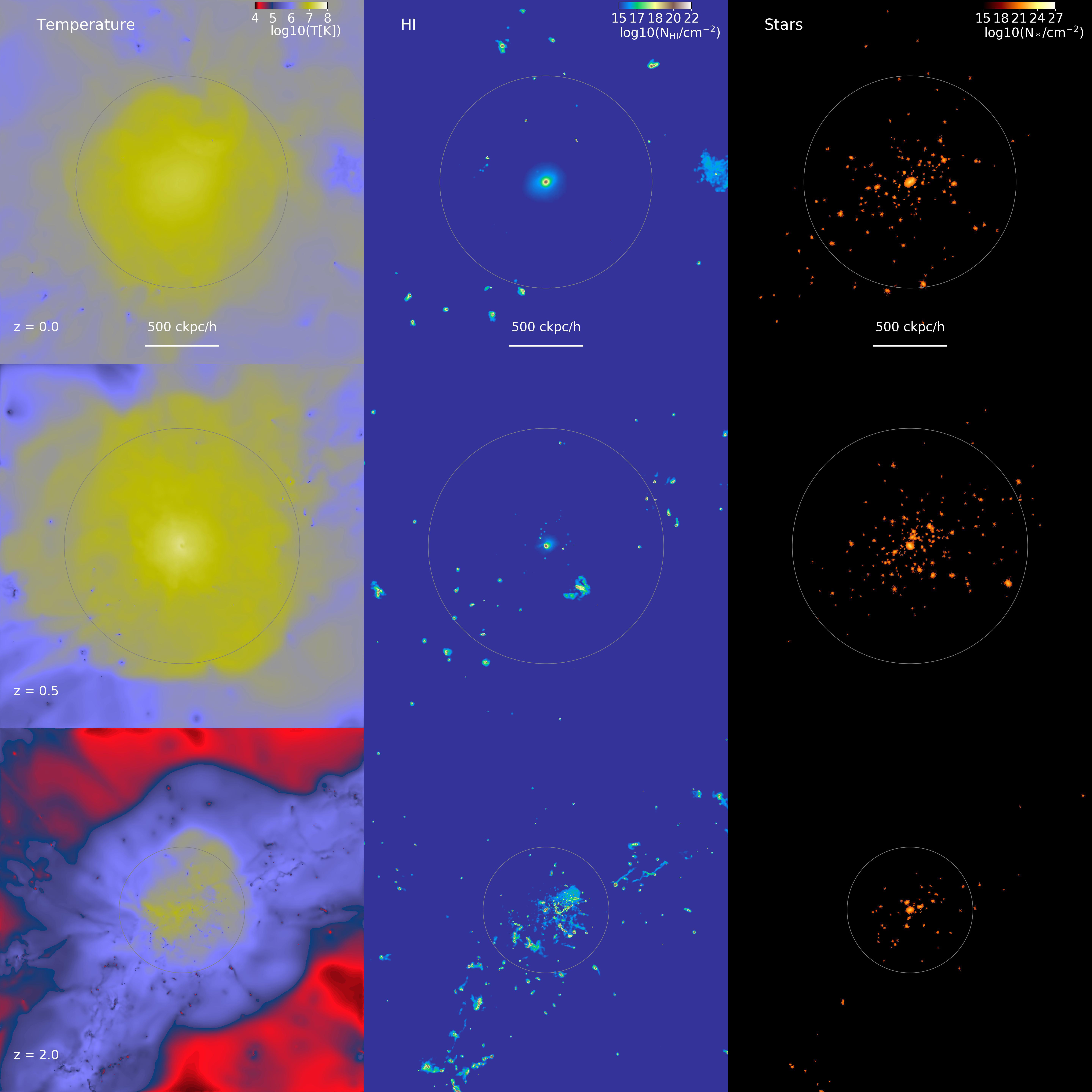}
\caption{
H10 at different redshift ({\it rows}. {\it Left} columns
show the temperature of the gas, {\it center} for the \HI~number
column densities and the {\it right} the star particles.
The panels are centered at the central galaxy.
Although \HI\ is present in the central galaxy
at $z=0$,  it does not form stars.
}
\label{fig_galaxycomponents}
\end{figure*}

Gas inflow is the main fuel for {\it in-situ} star formation in
galaxies.  The current understanding of galaxy evolution stipulates
that star formation-driven feedback ejects some of the infalling
gas to regulate stellar growth, while feedback associated with black
hole accretion prevents gas from reaching massive central galaxies
in haloes to quench galaxies \citep{Somerville-15r, Naab-17}.  In addition
to these processes, the immediate \citep[e.g.][]{Weinmann-06, Lancerna-18}
or extended \citep[e.g.][]{Kauffmann-13, Rafieferantsoa-18} environment
of galaxies can also affect their growth.  It has long been argued
theoretically that the cold gas content of galaxies is ram
pressure-stripped when galaxies move at high velocities through a
dense medium within a massive host halo \citep{Gunn-Gott-72}.
Furthermore, the extended diffuse gas around the galaxies can be
removed by the hot gaseous environment of dense structures such as
galaxy groups or clusters \citep{Larson-80}.  Observations of neutral
hydrogen being stripped from satellites in groups and clusters
directly support such claims \citep[e.g.][] {vanGorkom-03,
vanGorkom-04}.

Star formation of galaxies is apparently destined to end for galaxies
moving within dense environments.  An interesting question to examine
is the timescale on which such quenching of star formation happens,
for various types of galaxies, as this can provide insights into the
responsible physical mechanisms.  The enormous amount of data provided by the Sloan
Digital Sky Survey (SDSS) has permitted many studies of the effects
of galaxy environments on the star formation quenching timescales
in the nearby Universe. \citet{Kauffmann-04} used SDSS data to
analyse the correlation between star formation history and different
observed time indicators such as the $4000\AA$ break strength and
the Balmer-absorption index H$\delta_A$, and found no dependence
with environment, indicating a star formation timescale not less
than $1$ Gyr.  \citet{Peng-10} used SDSS with zCOSMOS \citep{Lilly-07}
to disentangle two different quenching mechanisms.  First, the {\it
environment quenching} --star formation quenching related to the
location of the galaxies -- is less of a function of redshift out
to $z\sim1$, which they argued to be a consequence of the formation
of large-scale structure that leads to the end of star formation
of over half of the satellite galaxies. Second, {\it mass quenching}
-- star formation quenching tied to the stellar mass of galaxies
-- was found to vary on a shorter timescale and is proportional to
the star formation rate of the galaxies.  \citet{Hirschmann-14}
estimated a timescale of $\sim 5$ Gyr for nearby low mass satellite
galaxies seen within SDSS, suggesting a very long quenching time.
Recently, \citet{Fossati-17} extended these results to higher
redshifts, using galaxies from the five CANDELS/3D-{\it HST} fields
\citep{Grogin-11, Koekemoer-11, Brammer-12} to estimate a quenching
timescale of $2-5$ Gyr out to intermediate redshifts.  They found
that the quenching processes were fairly independent of the host
mass, but were correlated with galaxy stellar masses and redshifts
such that smaller galaxies at lower redshift have longer quenching
time. These results, they argued, suggest that their galaxies stopped
forming stars due to fuel (gas) depletion. In addition, their
findings corroborate the {\it delayed-then-rapid} quenching scenario
that \citet{Wetzel-13} found locally using SDSS, in which galaxies
upon infall into a larger halo undergo a long delay phase where the
satellite behaves indistinguisably from a central galaxy, and a
short phase ($\la 1$ Gyr) where the star formation rate drops below
their detectability threshold.

These results outline a scenario where satellite galaxies retain
their gas reservoirs for a substantial period of time after infall
into a larger halo, but then rapidly lose both their gas and star
formation.  However, the gas depletion aspect of this is not directly
measured, but rather inferred from the behaviour of the satellite population.
The relative depletion times of the gas versus the star formation
rate can provide insights into which physical processes are
responsible.  For instance, rapid removal of extended neutral gas
while continuing star formation may indicate ram pressure stripping driving
the quenching of satellites, while concurrent reductions in gas
content and star formation may be more indicative of gas starvation.
Upcoming observations of \HI\ gas using telescopes such as the
Square Kilometre Array (SKA) precursor MeerKAT will be able to test
these scenarios directly.  It is thus timely to theoretically
investigate the co-evolution of gas and stars within haloes, both
in order to test whether current models reproduce the observed
behaviours, and also to make predictions for upcoming multi-wavelength
observations.

Much work has previously been done on satellite quenching from the
theoretical side.  \citet{McGee-09} studied accretion history with
a semi-analytic galaxy sample covering a wide range of environment,
{\it i.e.} from groups to clusters. Their sample showed that cluster
galaxies typically originate from smaller group galaxies, and as a
consequence the environmental effect on the galaxies lasted $>2$
Gyr.  Their finding suggests that at $z>1.5$, galaxies should
experience at most mild environment processes.  \citet{Simha-09}
used an SPH cosmological simulation and found a decreasing but
continuous gas accretion of the satellite galaxies.  They showed
that the gas depletion happens within $\sim 1-1.5$ Gyr timescale.
These timescales were already longer than those typically assumed
in semi-analytic models at that time, which \citep{Weinmann-06}
showed resulted in an overprediction of red satellite galaxies
compared with observations.

\citet{DeLucia-12} used a semi-analytic galaxy formation model based
on the Millennium Simulation to conclude that the majority of small
galaxies are pre-processed satellite, {\it i.e.} they were previously
located in smaller groups prior to their current one. By accounting
for such an origin, their model predicted a relatively long timescale
for satellite galaxies to lose their gas and halt their star formation
($\sim 6$ Gyr).  \citet{Wetzel-15} used high-resolution Local Group
simulations from the Exploiting the Local Volume in Simulations
(ELVIS) project, and found a rapid environmental quenching timescale,
shorter than $\lesssim 2$ Gyr for low-mass ($\sim10^8$ M$_\odot$)
satellite galaxies, with an additional $1-2$ Gyr when including the
pre-processing event.  The quenching timescale positively correlates
with the satellite stellar mass up to $\sim10^9$ M$_\odot$ up to
very large values of close to 10~Gyr, before decreasing for more
massive satellites down to $\sim 5$~Gyr.  Hence N-body plus
semi-analytic models generally require a quite long quenching
timescale in order to match observations. It is worth metioning
the Empirical ModEl for the foRmation of GalaxiEs
\citep[EMERGE,][]{Moster-18} which uses the delayed-then-rapid quenching
model to successfully fit the cosmic evolution of galaxy populations.

Our previous work in \citet{Rafieferantsoa-15} examined satellite
quenching using cosmological entropy-conserving Smoothed Particle
Hydrodynamics (SPH) simulations with a simple model for quenching
massive galaxies via gas ejection.  We found a halo mass dependent
timescale, where galaxies in more massive structures lose their gas
and shut off star formation faster than those living in less massive
host: $\sim1$ and $\sim2.5$ Gyr {\it e}-folding times respectively.
Qualitatively, this agrees with the semi-analytic model results,
but overall the timescales are shorter.  The range of predicted
quenching times suggest that the timescales depend on a variety of
factors, including physical ones such as halo mass and redshift,
and perhaps numerical ones such as resolution and hydrodynamic
methodology~\citep{Agertz-07}.  It would clearly be an improvement
to use cosmologically-situated gas dynamical simulations at high
resolution, including a more physically motivated quenching model
with more accurate hydrodynamics.

In this paper, we aim to do this using galaxies from high-resolution
cosmological zoom simulations of 2 galaxy groups with virial masses $>10^{13}M_\odot$
using the \giz\ code that better handles surface
instabilities~\citep{Hopkins-15}, and including a halo heating model
for radio mode quenching~\citep{Bower-06,Croton-06,Gabor-Dave-15}.
These zooms are extracted from the \muf\ cosmological
simulation~\citep{Dave-16} and re-run with nearly identical feedback
physics, but at $64\times$ better mass resolution.  We track
satellites falling into the primary groups in order to directly
quantify the duration from infall until when the galaxies lose their
gas and stop forming stars.  We quantify both the delay time when
satellites still behave similary to centrals of the same mass, and the
fading time over which satellites suddenly transition from being
gas-rich and star-forming to gas-poor and quenched.  The timescales
for gas evacuation and quenching are comparable.  The delay time
is generally $\sim 1-2$~Gyr, and drops rapidly with the infall mass
of the satellite, while decreasing modestly to higher redshifts.
In all cases the fading time is much smaller than the delay time
(typically $\la 0.2$~Gyr), and it is somewhat longer for more massive
infalling objects.  Our quenching timescale thus follows a {\it
delayed-then-rapid} scenario inferred from observations~\citep{Wetzel-13},
showing that this is a natural consequence of the interplay between
hydrodynamics and feedback physics within galaxy groups.

\S\ref{sim} briefly reviews the \muf\ simulation model used for
this work, with the description of the method used to get our galaxy
sample and find progenitors.  In \S\ref{gal_prop}, we examine the
distribution of galaxy properties used in this work, particularly
gas content and star formation histories.  In \S\ref{timescale},
we quantify the depletion timescale of \HI\ and SFR while \S\ref{gas-SFR}
and \S\ref{gas-Halo} show the relation between gas and SFR, and gas
and halo mass, respectively.  We summarize our findings in
\S\ref{conclusion}.

%\begin{figure*}
%%box_visual_dm.py
%\includegraphics[width=6in]{2D_135.jpg}
%\caption{stuff}
%\label{F:2D_dm}
%\end{figure*}

%---------------------------------------------------------------------------------------------------------------------------
\section{Simulations}\label{sim}
\subsection{Galaxy formation model}\label{model}
\indent

To analyse the evolution of galaxy \HI\ content, we use the \muf\
galaxy formation models fully described in \citet{Dave-16}.  The
same model was used in \citet{Rafieferantsoa-18} with minor
differences. This section briefly describes the main prescriptions
in \muf\ necessary for galaxy formation as well as the changes we
made to fit the context of this work.

First is the star formation rate (SFR) prescription that is based
on the molecular content of the gas particles. It follows a power
law scaling, namely the \citet{Schmidt-59} relation: \begin{equation}
\label{eq:sfr} \centering \mathrm{SFR} = \epsilon {\rho f_\mathrm{H2}\over
t_\mathrm{d}}, \end{equation} where $\epsilon = 0.02$ is the star
formation efficiency \citep{Kennicutt-98}, $f_\mathrm{H2}$ is the
molecular hydrogen fraction in the gas volume element and $t_\mathrm{d}$
is the dynamical time of the gas with density $\rho$.  The star formation
prescription is only applied to gas particles above a number density
threshold of $\mathrm{n_{thresh}} = 0.13~\mathrm{cm^{-3}}$. Such
high density is only reached when the gas particles undergo radiative
cooling.  \muf\ uses the {\sc Grackle 2.1}
library\footnote{\url{https://grackle.readthedocs.io/en/grackle-2.1/genindex.html}}
for cooling, which accounts for H/He-elements and metal-line cooling.
During the simulation, a uniform background metagalactic radiation
field is assumed following \citet{Faucher-Giguere-10}.

\begin{figure}
%MassEvolution.py
%\includegraphics{mass_evolution.pdf}
\includegraphics{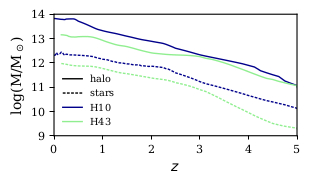}
\caption{
Total masses of the dark matter (full lines) and star (dashed lines) particles
{\it vs} redshift $z$ of H10 (dark blue) and H43 (light green) at the refined resolution.
}
\label{fig_mass_evolution}
\end{figure}

\begin{figure*}
%track_galaxies_HI.py
%\includegraphics[width=6in]{HALO_10_progen_track_HI.pdf}
%\includegraphics[width=6in]{HALO_10_progen_track_SFR.pdf}
\includegraphics[width=6in]{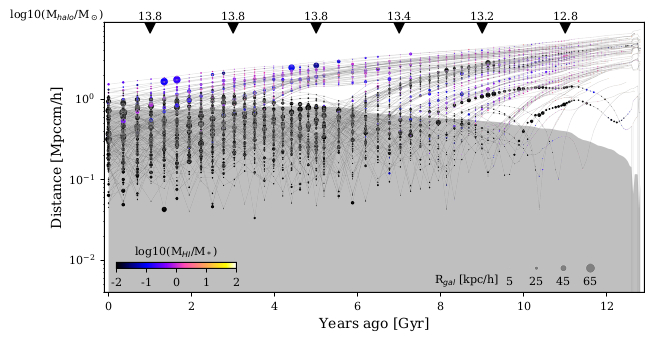}
\includegraphics[width=6in]{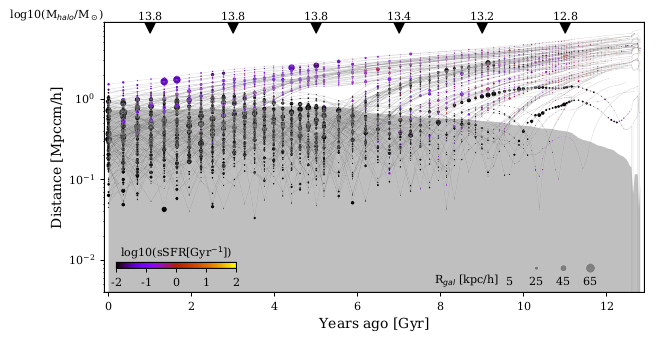}
\caption{
\HI\ (top) and SFR (bottom) evolution of galaxies in the H10 zoom
run. Each line shows the path of a
galaxy relative to its distance to the center of main halo.
The galaxies shown here are the galaxy members of the main halo
at $z=0$.  The shaded area is the region inside the virial radii
of the host. The evolution of the mass of the host is shown on the top
of the panel. 
%{\bf MIKA: Maybe down on the MHI/M* colorbar scale,
%so it goes from like 0.5 to -2.  Otherwise there is very little dynamic
%range to see evolution.  Meanwhile, for the sSFR, the trouble is that this
%quantity evolves very fast anyways.  So instead of sSFR, maybe show
%$\Delta$sSFR which is the difference relative to the main sequence (for all
%star-forming galaxies) at that redshift.  This should more clearly show quenching 
%within the virial radius, relative to the stuff outside.  Finally, you
%might want a colorbar that goes from blue to red to make things more visually
%evident.}
}
\label{fig_HIev10}
\end{figure*}

Galactic winds driven by supernovae and young star radiation pressure
are also accounted for in our model. Instead of modeling the detailed
physics that is beyond our resolution capabilities, \muf\ uses
scaling relations taken from the Feedback In Realistic Environment
\citep[FIRE,][]{Hopkins-14} simulations analysed in \citet{Muratov-15}
that model their combined effects into a mass outflow rate and
velocity. The mass loading factor $\eta$ and 
outflowing speed $v_w$ are thus taken to be:
 \begin{equation}\label{wind}
 \eta = 3.55\left(\frac{M_*}{10^{10}\msolar}\right)^{-0.35}
\hspace{0.25cm}\mathrm{and}\hspace{0.25cm}v_w = 2 v_c \left(\frac{v_c}{200\kms}\right)^{0.12},
 \end{equation}
where $M_*$ and $v_c$ are the stellar mass and circular velocity of the
galaxy which the gas volume element belongs to; we determine these quantities
using an on-the-fly friends-of-friends galaxy finder, and get the circular
velocity using the observed baryonic Tully-Fisher relation.  To eject gas
elements, we apply a velocity kick perpendicular to the local cross product
of the velocity and acceleration.

\begin{table*}%[t]
\begin{center}
%\resizebox{1\paperwidth}{!}{\begin{minipage}{\textwidth}
\caption[Table caption text]{Zoom simulated haloes.}
\begin{tabular}{c|c|c|c|c|c}
%\cline{2-5}
%\rule{0pt}{4ex}
%&\multicolumn{1}{c}{{\bf Coarse}}
%   &\multicolumn{2}{c}{{\bf Refined}}\\
%\cline{2-5}
%\rule{0pt}{3ex}
%& DM mass($\msolar$)
%   & Gas mass($\msolar$) & DM mass($\msolar$)\\
%\cline{2-5	}
%\rule{0pt}{4ex}
%& $1.14\times10^8$
%  &$2.85\times10^5$&$1.5\times10^6$\\
%\hline
\hline
Name & Original $M_{\rm halo} (\msolar)$ & Refined $M_{\rm halo}(\msolar)$
       & SFR$_\mathrm{central}$($\msolar$/yr)  & \# of satellites & $\mathrm{M_{*,central}/M_{halo}}$\\
\hline
H10   ($z=0$)    & $5.578\times10^{13}$ & $6.625\times10^{13}$ &  $0$ &  $182$  & 0.02441\\      
%mcen/mhalo = 0.01157
\hline
H43  ($z=0.15$) & $1.463\times10^{13}$ &   $1.422\times10^{13}$ &  $96.634$  &  $20$   & 0.05538\\
%mcen/mhalo = 0.03852
\hline
\hline
\end{tabular}\\ 
%masses are in $\msolar$.
\label{table:name}
%\end{minipage} }
\end{center}
\end{table*}

To prevent the overproduction of stars in massive ellipticals, \muf\
uses a heuristic prescription based on the hot halo quenching
scenario in \citet{Gabor-Dave-15}, which follows from semi-analytic
work showing that ``radio mode" quenching can yield a reasonable
quenched galaxy population~\citep{Bower-06,Croton-06}.  The idea
is to stop further supply of gas inflow for galaxies within host
haloes above a threshold quenching mass of $M_q >
(0.96+0.48z)\times10^{12}\msolar$, which is taken from the analytic
equilibrium model of \citet{Mitra-15}. In \muf\ this is done by
heating all the gas to the virial temperature $T_{vir} = 9.52\times
10^7 M_h^{2/3}$ \citep{Voit-05} of the haloes, excluding gas that
is self-shielded, i.e. has an \HI\ fraction above 1\% after the
self-shielding prescription of \citet{Rahmati-13} is applied.  One
significant difference with respect to the implementation in \muf\
is that we only heat the gas particles within the inner $25\%$ of
the virial radius of the haloes, rather than the entire halo, in
order to mitigate the overproduction of quenched satellite galaxies
as found in \citet{Rafieferantsoa-18}.

\subsection{Zoom simulations}\label{simg}

The cosmological parameters values used in this work are $\Omega_m
= 0.3$, $\Omega_\Lambda = 0.7$, $\Omega_b = 0.048$, $H_0 = 68
\kmsmpc$, $\sigma_8 = 0.82$ and $n_s = 0.97$, consistent with
\citet{-16}.  Our parent simulation is a $50\hmpc$ on a side with
$512^3$, with dark matter particles only, having identical random
seeds to the $50\hmpc$ \muf\ simulation in \citet{Dave-16}.  The
simulation starts at $z=249$ with initial condition generated using
{\sc Music} \citep{Hahn-11}.  Once at $z=0$, the haloes are identified
using the friends-of-friends algorithm with a linking length of 2\%
of the mean inter-particle distance.

\begin{figure*}
%HALO_MF.py
\includegraphics{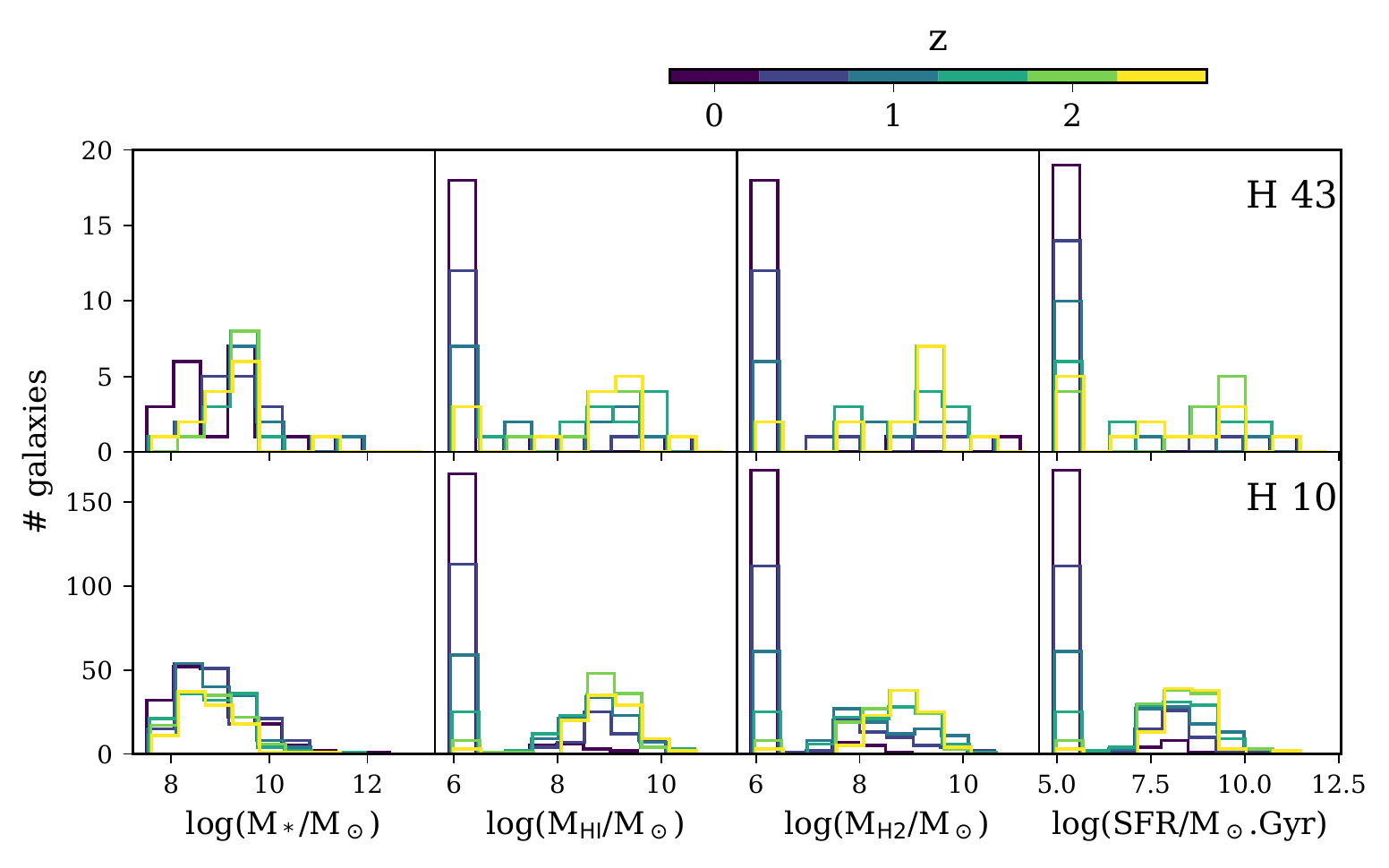}
\caption{
Evolution of the galaxy property distributions of our selected haloes.
{\it Upper} panels are for H43 and {\it lower} panels for H10.
Different colors represent the distributions at different redshift.
Each columns show distributions for different galaxy properties.
}
\label{fig_HMF}
\end{figure*}

\begin{figure*}
%SFR_Timescale.py
%\includegraphics[scale=0.85]{SFR_timescale.pdf}
%\includegraphics[scale=0.85]{HI_timescale.pdf}
\includegraphics[scale=0.85]{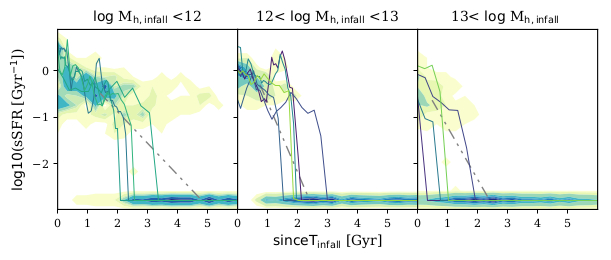}
\includegraphics[scale=0.85]{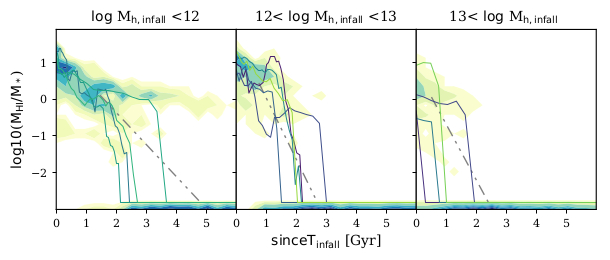}
\caption{
Specific star formation rate ({\it upper panels})  and \HI-richness ({\it lower panels})
of galaxies since they became satellite.
The mass ranges of the halo at infall (M$_\mathrm{h,infall}$)
are shown on top of the panels. Note that all the galaxies belong
to H10 at $z=0$. It is clear that galaxies still form stars
for $\geq 5$ Gyr when M$_\mathrm{h,infall} \leq 10^{12}$ M$_\odot$.
That time decreases down to $\sim 1$ Gyr for M$_\mathrm{h,infall}
> 10^{13}$ M$_\odot$. The gray dashed lines connect the median values of the
upper and lower contours.
The lines show tracks of representative galaxies in each
M$_\mathrm{h,infall}$ range.
}
\label{fig_SFRtimescale}
\end{figure*}

\begin{figure}
%TimescaleCombine.py
%SFR_DelayTimeCompare.py
%SFR_FadingTimeCompare.py
%\includegraphics{QuenchingTimeCompare.pdf}
\includegraphics{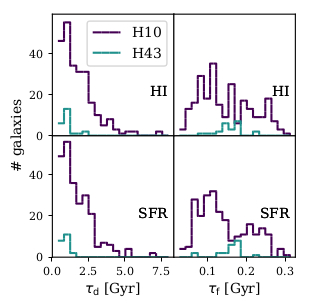}
\caption{Delay time ({\it left panels}) and fading time ({\it right panels})
distributions of the galaxies at $z\leq1$ quantified by the SFR ({\it lower panels})
and the \HI\ content ({\it upper panels}). Purple lines for H10
and blue lines for H43}
\label{QT}
\end{figure}

We apply the zoom-in technique for two carefully chosen haloes within
our parent box, in which we re-simulate these haloes with all physical models
at significantly higher resolution.  To investigate the quenching
of satellite galaxies, we focus on the mass range found in
\citet{Rafieferantsoa-18} where satellite quenching effects were
most evident, namely $M_h\sim 10^{13}M_\odot$.  Hence we select
halo 10 (H10, hereafter) and halo
43 (H43, hereafter), having $z=0$ masses
of $5.5\times 10^{13}M_\odot$ and $1.7\times 10^{13}M_\odot$ at coarse
resolutions. This is well above our quenching mass scale $M_q\sim10^{12}\msolar$,
but nonetheless we find that the smaller halo has a central galaxy
that is still star forming, owing to remaining cold gas in the halo
and infall of gas-rich satellites.  The larger halo has a quenched
central.  Our choice of haloes is further based on their evolutionary
growth history, namely that these haloes did not have any major
fly-by or merger during their evolution. This is to ensure that the
haloes are evolved in a self-contained way and did not change their
physical properties by interacting with other structures of similar
size.

Our two haloes are re-simulated with dark matter particles $64$ times
lower mass than those from the primary box within the zoom region,
plus a surrounding region with $8\times$ better resolution. The
$64\times$ refined dark matter particles are further split into gas
volume elements and dark matter particles based on our adopted
baryon fraction (see \S\ref{simg}).  When resimulating, the center
of mass of the highly resolved particles is shifted to the centre
of the cubical low-resolution volume.  Properties of the haloes
presented in this work are sumarized in Table \ref{table:name}.
At the refined resolution, the H10 actually becomes slightly
larger than in at the parent resolution, up to $6.625\times 10^{13}M_\odot$ ($z=0$),
while H43 drops slightly to $1.422\times 10^{13}M_\odot$ ($z=0.15$).  The former
has no central star formation, while the latter is forming stars
vigorously at $\sim 100 M_\odot/$yr.  The former also has far more
satellites, 180 vs. 20, and hence the statistics of satellites in
this paper is dominated by this larger H10 halo.

For illustration, Figure \ref{fig_galaxycomponents} shows H10 at
$z=0,0.5,2$ (top to bottom), with the left column showing the
projected gas temperature, the middle column showing \HI\ column
density, and the right column showing the stellar mass surface
density.  The circles indicate the sphere enclosing $200\times$ the
critical density at those epochs.  At $z=2$ (10~Gyr ago) it is
evident that \HI\ is distributed with the stars throughout the
halo, and even shows some diffuse features outside of galaxies.  By
$z=0.5$ (5~Gyr ago) there is a clearer segregation of \HI\ towards
the outskirts of the halo, though there remain some \HI-rich
galaxies within the halo.  By $z=0$, very little \HI\ is left in
the halo, as the hot $\sim 10^7$K gas has grown to encompass the
halo and beyond.  This visually demonstrates the strong impact that
environment and halo quenching have on the gas content of satellites.

Figure \ref{fig_mass_evolution} illustrates the growth history in
the total dark matter (full lines) and stellar (dashed lines)
components in the H10 (dark blue) and H43
(light green) high resolution runs. These show that the haloes masses grow roughly
in concert as expected (solid lines), but that the stellar mass
(dashed lines) of the smaller halo grows more slowly until $z\sim
3$, and more rapidly at later epochs.  In the end, the total stellar
to halo mass ratio for the small group ends up significantly larger.
These two haloes thus span the interesting mass regime where the
smaller one is significantly retarded by star formation feedback
at early epochs, while the more massive has an early-weighted growth
history more typical of quenched massive galaxies today.  These
growth histories may help explain why H10 is fully quenched today,
but H43 remains star-forming. One way to look at this would be to show
the baryon conversion efficiencies byt we leave such investigation of growth
rates for future work, with larger samples.

Similar to other \muf\ papers, we use {\sc
Skid}\footnote{\url{http://www-hpcc.astro.washington.edu/tools/skid.html}}
galaxy finder and the properties of galaxies and haloes are obtained
from {\sc Caesar}\footnote{\url{https://bitbucket.org/laskalam/caesar}},
which uses the \texttt{yt} simulation analysis suite as a backend.
See \citet{Rafieferantsoa-18} for details on the calculation of those
properties including M$_\mathrm{halo}$, M$_*$, M$_\mathrm{HI}$, M$_\mathrm{H2}$
and SFR of the galaxies.
%--------------------------------------------------------------------------------------

%--------------------------------------
%==================================
\section{Evolution of gas and stellar properties} \label{gal_prop}

In this work we are primarily interested in the gas content and
star formation history of the galaxies in our host haloes.  Figure
\ref{fig_HMF} shows the distributions of stellar mass ({\it left}
column), the \HI~mass ($2^\mathrm{nd}$ column), the H$_2$
mass ($3^\mathrm{rd}$ column) and the star
formation rate ({\it right} column) of the galaxies in H10
({\it lower} panels) and in H43 ({\it upper} panels).
The different coloured curves show the evolution from $z\approx
3\rightarrow 0$ (yellow to violet).  Galaxies are given lower-limit
values once their \HI, H$_2$, or SFR goes below the x-axes limit
of our figure.

In the $M_*$ panel, one sees a buildup of smaller accreted satellites
within the halo as time passes, while the most massive galaxy grows
steadily in time.  The gas content is higher at intermediate
redshifts, and then the population of gas-free satellites builds
up at low redshifts.  There is a lot of similarity in the \HI,
H$_2$, and SFR distributions, showing that these quantities seem
to broadly evolve together, even within a galaxy halo.  The H43
halo (lower row) shows qualitatively similar trends, though the
galaxies themselves tend to be less massive, and there are many
fewer satellites (note the $y$-axis scale).  Given these distributions
of galaxy properties in our haloes, we can look at the evolution of
\HI~and H$_2$ content and their timescales for consumption.

With our simulated outputs spaced by $\sim300$ Myr at $z\sim 0$ and
shorter intervals at higher redshift, we can track individual
galaxies based on their stellar masses. We match the galaxies between
two successive snapshots and identify the galaxies with the maximum
number of similar star particles. An issue with this method is that
the progenitor of a smaller galaxy flying by a bigger galaxy would
have its progenitor always then be the bigger galaxy.  To avoid
this, we further match the galaxies between snapshots at two separated
time intervals, in order to avoid mischaracterising fly-bys as
mergers.

Figure \ref{fig_HIev10} shows the distances to the progenitors of
galaxies present in the host halo (H10) at $z=0$.
The size of the circles show the total baryonic mass of the galaxy
(gas+stars), and the colour of the circles show the properties of
the galaxies as indicated in the colour bar -- \HI\ in the top
panel, and SFR in the bottom panel.  The shaded region represents
$R_{200}$ at any given epoch.  The groups build up hierarchically.
In this work, we are only concerned about the progenitors of the
galaxies present in the main halo at $z=0$.  We can clearly see
that when galaxies cross the virial radius of the halo of interest,
their \HI-richness (or sSFR) is considerably depleted and the only
growth of galaxies is via dry mergers \citep[{\textit{e.g.}}][]{Oser-10}.
The following sections try to quantify the timescale required for such events.
Througout this work, we use H10 and H43 data
except for the latest redshift where we only use H10 because
H43 was stopped at $z\geq0.15$.

\section{satellite quenching timescales}

\label{timescale}
\begin{figure*}
%SFR_DelayTime.py
%HI_DelayTime.py
%\includegraphics{SFRDelayTime.pdf}
%\includegraphics{HIDelayTime.pdf}
\includegraphics{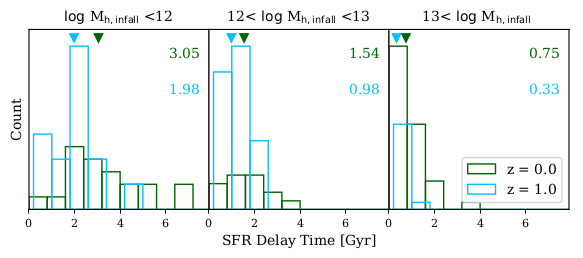}
\includegraphics{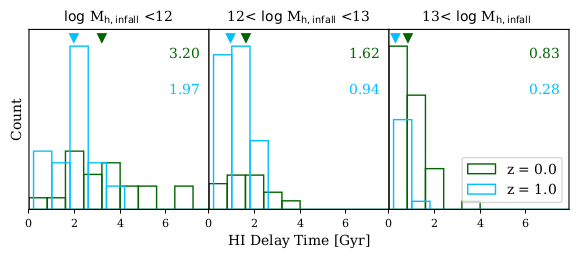}
\caption{{\it Upper} ({\it Lower}) panels show the sSFR
(\HI-richness) delay time distributions. {\it Left} panels are
for the galaxies first falling in $\log\mathrm{M_{h,infall}} <12$,
the {\it Middle} panels for $12<\log\mathrm{M_{h,infall}} <13$
and the {\it Right} panels for $13<\log\mathrm{M_{h,infall}}$.
The downwards triangles on the top of the panels point to
the mean values of the delay times where the corresponding
values are shown on the top rights (color coded).}
\label{fig_SFRDelayTime}
\end{figure*}

Cold gas within galaxies is diminished because of environmental effects
within the virialised regions such as stripping (ram pressure and
tidal) and the lack of gas inflow (or starvation), along with the consumption
of gas in the interstellar medium via star formation and outflows.  Estimated
observed timescales for quenching are $t_q$ of $5-7$ Gyr \citep{Wetzel-13}
or $1-2$ Gyr \citep{Oman-Hudson-16}, depending on the galaxy masses
and the environment they live in: the more massive the galaxies or
the haloes, the shorter the timescales.  Here we quantify these
timescales using our zoom simulations to emphasise on the differences
between the gradual starvation of the galaxies by lack of inflowing
cold gas, and the rapid drop of star formation rate due to the
remaining molecular content being used up.

%{\bf MIKA: I don't think starvation is gradual, otherwise we
%would already see differences between sats and centrals during the
%delay phase, but we don't. ROMEEL: Why not? But isn't this what we see in figure 5.
%First, sSFR decrease slowly (gradual), then suddenly drops? The difference between
%sats and cens might be difficult to see, but we show that the difference is there between
%sats and isolated cens in our previous paper.}

We define the starting time $t_0$ of the gas depletion and star
formation quenching to be when a galaxy falls into a more massive
halo and becomes satellite.  We only include galaxies that
were always central prior to $t_0$.  \citet{Fossati-17} used the
same definition of $t_0$, and they found a 2-5 Gyr timescale for
quenching the satellites.  They found that the fraction of quenched
satellites is higher at lower redshift and higher halo masses.
Surprisingly, they found no dependence of the passive central galaxy
fraction with halo masses (see their fig. 17).

Figure \ref{fig_SFRtimescale} shows the 2D distributions in time
after $t_0$ versus the sSFR (top panel) and the \HI\ fraction
(bottom) for satellites in H10.  The left panels
are for galaxies falling in primaries with M$_\mathrm{halo} <
10^{12}$ M$_\odot$, the middle panels show $10^{12}$ M$_\odot <$
M$_\mathrm{halo} < 10^{13}$ M$_\odot$, and the right panels show
M$_\mathrm{halo} > 10^{13}$ M$_\odot$.  The dashed lines go from
the median values of the upper contours to the median values of the
lower countours.

We define the quenching timescales ($t_q$) to be the time since
$t_0$ when the galaxy last reaches \HI-richness$_\mathrm{~lim} <
10^{-2}$, or sSFR$_\mathrm{~lim} < 10^{-2}$ Gyr$^{-1}$; thus we
have a gas quenching timescale and a SFR quenching timescale.  The
dashed lines in Figure \ref{fig_SFRtimescale} suggest a shorter
$t_q$ for higher infall halo masses.  This is in contrast to
\citet{Fossati-17} who did not find any such dependence in their
observed sample. We speculate this to originate from the difference
between the simulated and observed galaxy properties. For instance,
the halo mass of the simulated galaxies are computed using the
friends-of-friends technique directly on the 3D-position of the
galaxies, whereas a matching technique between observed and mock
galaxies is used for the 3D-HST sample in \citet{Fossati-17}. The
uncertainty in redshift space of $\pm 0.1$ between the observed
galaxies and the the mock catalogue might impact their finding on
the environment dependence of the quenching timescale. In addition,
the stellar mass threshold of their sample becomes as high as
$>10^{10}\msolar$ for $z>1$, affecting the galaxy number density
and the matching technique thereafter.

\subsection{Computing delayed-then-rapid quenching} 

We further split $t_q$ into a {\it Delay Time} $\tau_\mathrm{d}$ during which
the satellite properties are similar to a central galaxy of the
same mass, and a {\it Fading Time} $\tau_\mathrm{f}$ during which strong depletion
happens and the \HI-richness or sSFR reaches the threshold limit.
We note that those definitions are similar to those used in \citet{Fossati-17}.

We compute $\tau_\mathrm{d}$ in our simulations as the time between
$t_0$ and the earliest snapshot before the instantaneous rapid drop
of sSFR or \HI~richness occured.  $\tau_\mathrm{f}$ is then the remainder of
the time to quench ($t_q=\tau_\mathrm{d}+\tau_\mathrm{f}$), i.e.  the duration from the
beginning of the rapid drop to the next time the quantity is below
the threshold.  The fading time typically happens within two
successive snapshots in our simulation, that is, on timescales less
than a couple hundred Myr.  In that case we interpolate the value
based on an exponential fit, but these numbers may be uncertain.

For $z\leq1$, we show in Figure \ref{QT} the histograms of the
delay time and the fading time for satellites in H10
(purple) and H43 (blue). The distributions of the
delay times (left panels) show similar trends between the two haloes,
indicating a power law distribution of delay times. In contrast, the
fading times (right panels) are less similar, {\it i.e} $\tau_\mathrm{f}$
from H10 tend to show double peak distributions
whereas those from H43 with only single peaks. The
difference is less conclusive due the very small number of galaxies
in H43.

\subsection{Delay Time} 
\label{delaytime}

Most of the quenching time is spent in the delay phase where satellite
galaxies evolve similarly to central galaxies. \citet{Wetzel-13}
and \citet{Fossati-17} found $\tau_\mathrm{d} > 80\%$ the total
quenching time.  Figure \ref{fig_SFRDelayTime} shows the distribution
of delay times for the sSFR (upper) and the \HI\ fraction (lower)
of galaxies.  The downward triangles at the top of the panels show
the  mean values of the delay times at different redshift: $z=0$
(dark green), $z=1$ (light blue).  Left to right panels
show the results for increasing halo mass at first infall.  The
distributions span a wide range of values especially for $z=0$. The
mean values differences between the two different redshift are
higher for lower halo mass at first infall. The different timescales
at different epochs hint a star formation efficiency difference
that we will explore in \S\ref{gas-SFR} and \S\ref{gas-Halo}.  SFR
and \HI\ delay times show similar trend in their distributions.

\begin{figure}
%DelayTime_Mstar.py
%\includegraphics[scale=0.8]{DelayTime_Mstar.pdf}
\includegraphics[scale=0.8]{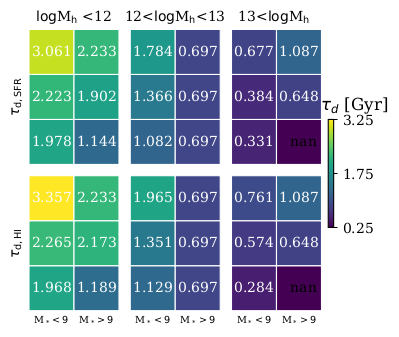}
\caption{
Delay time $\tau_\mathrm{d}$ for sSFR ({\it upper} panels) and \HI-richness
({\it lower} panels). {\it Left} panels are for galaxies becoming satellite
in $\mathrm{M_{h,infall}}<10^{12} M_\odot$, {\it center} panels for $10^{12} \leq\mathrm{M_{h,infall}}<10^{13} M_\odot$
and {\it right} panels for $10^{13} M_\odot\leq\mathrm{M_{h,infall}}$. Each panel is further
divided into 6 area: 2 columns for stellar mass bins at infall
($\mathrm{M_{*,infall}}<10^9 M_\odot$ {\it left} and
$\mathrm{M_{*,infall}}\geq10^9 M_\odot$ {\it right}) and 3 rows 
for different redshifts ({\it upper} for $z=0$, {\it middle} for $z=0.5$ and {\it lower}
for $z=1$). See Figure \ref{fig_SFRDelayTime} for the distribution of galaxies in terms of $\tau_\mathrm{d}$.
}
\label{fig_DelayTime_Mstar}
\end{figure}

To further analyse these trends, we also bin our data based on the galaxy stellar
masses and look at the evolution of the delay timescales at different
redshift.  Figure \ref{fig_DelayTime_Mstar} shows the sSFR and
\HI-richness mean delay times $\tau_\mathrm{d}$: {\it upper} and
{\it lower} panels respectively. {\it Left} to {\it right} panels
show the results for increasing halo masses within which the galaxies
first fall in. Each panel is divided into 6 sections: two columns for
binned stellar masses of the infalling galaxies at $t_0$ ({\it
left}: M$_\mathrm{*,infall} < 10^9\mathrm{M_\odot}$ and {\it right}:
M$_\mathrm{*,infall} \geq 10^9\mathrm{M_\odot}$) and three rows for
different redshift ({\it upper} for $z=0$, {\it mid} for $z=0.5$
and {\it lower} for $z=1$). The numbers on the figure show the
average $\tau_\mathrm{d}$.  We note that at higher redshift, we
only track the galaxies present in the main halo progenitors at
that redshift, {\it i.e} we do not include the progenitors of
galaxies that are present in the $z=0$ main halo but did not cross
the virial radius by that given redshift.

The mean of the delay times is the longest for the least massive
host halo at $z=0$ ($\sim 3$ Gyr), and decreases down to $\sim1$
Gyr by $z\sim1$.  The galaxies falling in the most massive halo of
infall take the shortest delay time: $\sim1.5$ Gyr faster than those
falling in M$_\mathrm{h,infall} \leq 10^{12}$ M$_\odot$ ({\it i.e.}
$\leq 1$ Gyr).  Less massive satellite galaxies have longer delay
time (1 Gyr more) than their massive counterparts, except for those
falling in the most massive structure where this scenario is flipped.
\HI-richness and sSFR delay times show similar behaviour, which 
we explore in more detail later.

\subsection{Fading Time} \label{fadingtime}

In the delayed-then-rapid scenario, the fading time $\tau_\mathrm{f}$
is expected to be much shorter than the delay time. \citet{Fossati-17}
computed the delay time by matching the star formation efficiency,
quantified by the main sequence parameterisation of \citet{Wisnioski-15},
between the 3D-HST sample and their mock catalogue.  They estimated
$\tau_\mathrm{f} = t_q - \tau_\mathrm{d} <0.6$ Gyr.

\begin{figure*}
%SFR_FadingTime.py
%HI_FadingTime.py
%\includegraphics{SFRFadingTime.pdf}
%\includegraphics{HIFadingTime.pdf}
\includegraphics{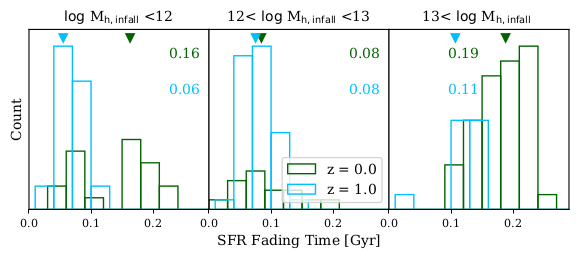}
\includegraphics{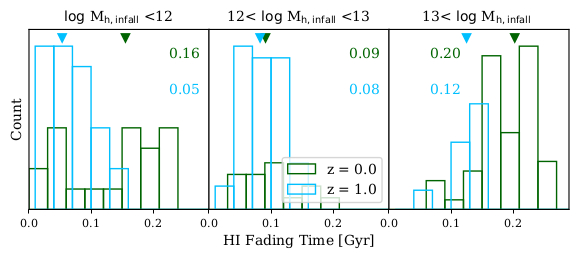}
\caption{Similar to Figure \ref{fig_SFRDelayTime} but showing for the
Fading Time $\tau_\mathrm{f}$.}
\label{fig_SFRFadingTime}
\end{figure*}

\begin{figure}
%FadingTime_Mstar.py
%\includegraphics[scale=0.8]{FadingTime_Mstar.pdf}
\includegraphics[scale=0.8]{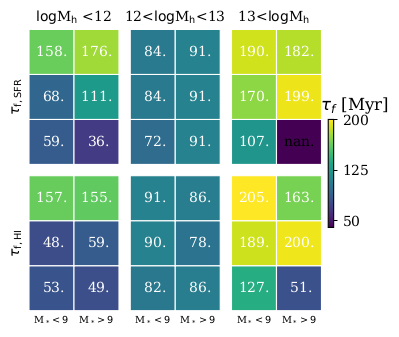}
\caption{
Similar to Figure \ref{fig_DelayTime_Mstar} but showing for the
Fading  time $\tau_\mathrm{f}$ binned by stellar mass.
}
\label{fig_FadingTime_Mstar}
\end{figure}

Similar to \S\ref{delaytime}, we also look at the timescale
distributions as shown in Figure \ref{fig_SFRFadingTime}. In this
case, the mean values (downward triangles) do not show clear patterns
with halo mass at first infall. With respect to time, we can
see that the longer timescale at lower redshift is still present.
Keep in mind that owing to the small values for $\tau_\mathrm{f}$ and the
spacing of our simulation snapshots, the values are less precise
and some trends may be washed out.

Figure \ref{fig_FadingTime_Mstar} shows the mean fading times
$\tau_\mathrm{f}$ of the \HI-richness ({\it lower} panels) and sSFR
({\it upper} panels).  All scenarios show mean fading time of $\sim
50-200$ Myr. Generally, lower redshift ({\it upper} parts of each
panel) infall requires longer $\tau_\mathrm{f}$ to suppress the
star formation and the neutral gas content compared to their
counterparts at higher redshift ({\it lower} parts of each panel).
Interestingly, galaxies falling into the most massive host tend to
require more time to completely consume their gas and stop star
formation, e.g. $\sim185$ Gyr for sSFR and $\sim200$ Myr for \HI-richness
for galaxies at $z=0$, which might be due to our limited sample and
requires further analysis.  The stellar mass of the galaxies does not
show any trend in the fading time, {\it i.e.} difference between
the {\it left} and {\it right} areas in each panel.

In summary, the delay time $\tau_\mathrm{d}$ is typically in the
range of $\sim 1-3$~Gyr, longer for lower stellar and halo masses
until the haloes exceed $10^{13}\msolar$. In the latter case, less
massive galaxies have $\tau_\mathrm{d}$ of $\sim400$ Myr lower than more
massive ones. Galaxies have longer $\tau_\mathrm{d}$ when they
become satellite more recently. The numbers are very similar for
both \HI\ and sSFR, which suggests that the SFR is being attenuated
owing to the removal (or stripping) of the extended gas reservoir. The fading
time $\tau_\mathrm{f}$, in contrast, is always very short, and shows
the opposite trends. Generally, galaxies falling into higher halo masses
have longer $\tau_\mathrm{f}$. The
case is less clear at $z=0$ but becomes more apparent at higher
redshift.  The difference for lower and higher galaxy stellar masses
is minor if nonexistent.

At face value, we find similar trends to \citet{Fossati-17} in
terms of stellar mass, but our timescale dependence with halo mass
was not seen in their 3D-HST sample, and our delay times are generally
shorter by a factor of two or so.  Assessing the significance of
these discrepancies likely requires performing a more careful
comparison where we create a mock sample with grism data and
photometry and select and analyse this in the same way; we leave
this for future work.  For now, we note that the broadly similar
trends as seen in observations both locally and intermediate redshifts
are encouraging, and we predict for future \HI\ surveys that the
delay and fading times should be similar for the gas and SFR, at
least in this halo mass range.

\section{Relation between gas content and Star formation}
\label{gas-SFR}
\begin{figure}
%GasSFR_HI.py
%\includegraphics{GasSFR_HIev.pdf}
\includegraphics{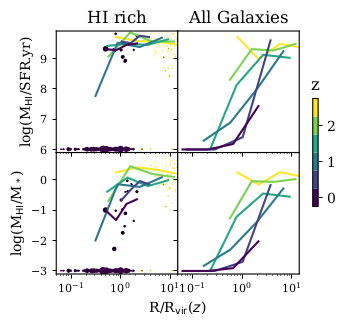}
\caption{
\HI-consumption time ({\it upper panels}, $\log(\mathrm{M_{HI}/SFR})$)
and \HI~richness ({\it lower panels}) of galaxies {\it vs} their radial
distances from the center of the host halo. The circles show the
galaxies at the lowest (purple) and highest (yellow) redshift
presented here.
The lines show the mean values, colour coded by the redshift.
Higher redshift galaxies are the progenitors of the low redshit galaxies.
The low redshift galaxies are only the members of the main halo.
Galaxy properties below $6$ (for \HI-consumption time) and $-3$
(for \HI-richness) are given those lower limit values 
to confine the figure. {\it Left} panels show the mean values
for only gas rich galaxies but {\it Right} panels show the mean
values for all galaxies.
}
\label{fig_HISFR}
\end{figure}

The cooling and coalescence of neutral hydrogen into dense molecular
clouds are important steps before the star formation. However, the
formation of stars can disrupt the collapse of cold gas and therefore
reduce later star formation.  To better understand the driver of
satellite quenching, here we quantify the star formation activity
given the amount of cold atomic and molecular gas.

\begin{figure}
%GasSFR_H2.py
%\includegraphics{GasSFR_H2ev.pdf}
\includegraphics{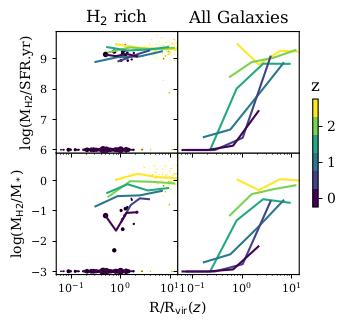}
\caption{
Similar to Figure \ref{fig_HISFR} except showing for H{$_2$} content.
}
\label{fig_H2SFR}
\end{figure}

{\it Left} panels of Figure \ref{fig_HISFR} show the evolution of
\HI-richness ({\it lower} panel) and \HI-consumption time
(M$_\mathrm{HI}$/SFR, {\it upper} panel) with respect to their
distance to the main and its progenitors center of masses.  Note
that we only follow all the progenitors of the galaxies present in
the $z=0$ main halo. The purple dots are the galaxies at $z=0$,
with the purple lines showing the mean values. Yellow dots are the
progenitors of the $z=0$ galaxies at $z=2.5$, with the yellow lines
the mean values. Lines with different colors are the mean values
at different redshift as shown in the colorbar. We did not show the
galaxies themselves to avoid cluttering of the figures. Galaxies
without cold gas are shown at the bottom of each panel but
are not included in the calculation of the mean quantities. This is
because we are only interested in those that still have \HI. The
virial radii used in the x-axes is for the progenitors of the main
halo not for the individual galaxies.

A first look, the {\it left} panels of Figure \ref{fig_HISFR}
indicates that our galaxy sample has higher gas content at higher
redshift, the galaxies lose their gas until today ({\it lower}
panel), and those crossing the virial radius contain less
\HI. The \HI-consumption time ({\it upper} panel) is barely a
function of redshift and distance to the center of the main halo
progenitors. This reinforces the idea of gas consumption via star
formation, and the quenching of the latter by depletion of the
former.  This scenario remains the same at different cosmic times.
We speculate that the dichotomy present in \HI-consumption time
being analogous to that in \HI-richness indicates the existence of
an \HI-richness threshold below which further star formation is unlikely
to happen without any extra supply of cold gas and above which the
star formation rate correlates directly with the \HI\ content, even
though the star formation is actually occuring within the H$_2$
gas.

Left panels of Figure \ref{fig_H2SFR} show the H$_2$-richness and
H$_2$-consumption time to further analyse the gas consumption.  The
relationship is cleaner compared to the previous one, mainly due
to our star formation prescription scaling with the H$_2$ fraction
($f_\mathrm{H2}$, see equation \ref{eq:sfr}). Again, the pattern
shows that galaxies contain less gas in molecular form at lower
redshift than at higher redshift ({\it lower} panel) and their star
formation rate proportionally declines with it ({\it upper} panel),
{\it i.e.} unchanged H$_2$-consumption time at different radial
distance and different redshift. The similarity between the
\HI-consumption time and the H$_2$-consumption time  suggests the
\HI\ to be a direct tracer of star formation rate above a certain
mass threshold. To test this, we can increase our group sample and
vary the gas element density thresholds for self-shielding and star
formation.  We plan to investigate on this in future work.

Statistically, we can do the same exercise but including all the
galaxies regardless of their gas content. In this case, we are
looking at the general trend rather than the process towards being
quenched as we argued previously.  Right panels of Figures
\ref{fig_HISFR} and \ref{fig_H2SFR} are similar to the left, except
that we now show the mean values for all the galaxies. The trend
remains the same at the highest redshift, but the growth of no-cold-gas
galaxies at lower redshift shifted down the mean values.

In short, the gas consumption timescale is neither a function of
the group-centric distance nor the cosmic time. This is seen in
both atomic and molecular hydrogen contents of the galaxies suggesting
that above a certain mass threshold, \HI\ alone can be a direct
indicator of star formation. We note that our star formation prescription
directly ties the H$_2$ content to star formation rate, while there is likely
to be a more stochastic correlation that is sensitive to local interstellar
medium conditions which are unresolvable in our cosmological simulations
\citep{Hu-16, Naab-17}. Hence the tightness of H$_2$ and SFR may be partly an
artifact of our star formation prescription, but the tightness with the
\HI\ and SFR is unlikely to be significantly impacted by this

\section{Relation between gas content and halo mass at infall}
\label{gas-Halo}

We showed previously that satellite galaxies have shorter delay
times when they fall into more massive hosts.  In that vein, we
look at the change of gas-richness and gas-consumption time depending
on the first halo of infall in this section.

\begin{figure}
%Gas_Minfall.py
%\includegraphics{GasSFR_Minfall.pdf}
\includegraphics{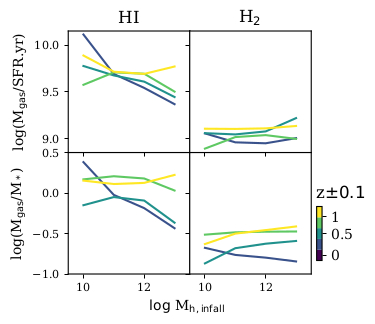}
\caption{
Gas-consumption time ({\it Upper}) and gas-fraction ({\it Lower})
of galaxies {\it vs} the halo mass they first fall in.
{\it Left} ({\it Right}) panels are for \HI~(H$_2$). We use a
binwidth of 0.2 around the redshift values (hence the $z\pm0.1$ on the colorbar).
}
\label{fig_GasSFR_Minfall}
\end{figure}

Figure \ref{fig_GasSFR_Minfall} shows the gas-richness ({\it lower}
panels) and gas-consumption time ({\it upper} panels) of galaxies
{\it vs} the mass of the haloes they fall in.  The left panels show
for \HI~content whereas the right panels for H$_2$ content.  Generally,
the decrease of \HI-richness is faster when it falls in more massive
haloes ({\it lower} panel) similar to what was found in
\citet{Rafieferantsoa-15}. However, such attributes are only true
at lower redshift (purple, $z\sim 0$) but is not seen at the higher
redshift explored here (yellow, $z\sim1$). The {\it upper} panel
shows that galaxies form star less efficiently early on for a given
halo mass of infall:{\it i.e.} yellow $>$ green $>$ blue.  The trend
is very robust at higher halo mass of infall while at lower host
halo masses, the trend becomes less coherent.  This is due to the
decreasing number of galaxies above our \HI~richness threshold at
later time as infall in less massive host mostly happened earlier
than infall in more massive ones.  The \HI-consumption time is
independent of halo mass at infall at higher redshift but anti-correlates
strongly at lower redshift.

In terms of molecular content, H$_2$-richness is less of a function
of halo mass of infall than \HI-richness and the correlation is
barely present if not missing at any $z\lesssim1$.  H$_2$-richness
at high halo mass of infall is higher at higher redshift, while it
is less conclusive at the lowest masses.  H$_2$-consumption time
is flat with respect to the halo mass of infall.  We barely see any
evidence of consumption time difference at different redshift that
was relatively apparent with \HI-consumption time at higher halo
mass of infall.

\section{Conclusion}
\label{conclusion}

We use two zoomed-in groups of galaxies to study the timescales
at which the \HI~content and the sSFR deplete when the galaxies
become satellite.  We determine the timescale for quenching $t_q$,
and use observationally-motivated definitions to partition this
into a delay time $\tau_\mathrm{d}$ during which there is a gradual drop in
star formation rate and gas content, and a fading time $\tau_\mathrm{f}$
when the galaxy becomes quenched.  We compare the results for
$\tau_\mathrm{d}$ and $\tau_\mathrm{f}$ for star formation quenching, and for \HI\
removal.  We also quantify the efficiency of gas to form
stars as well as the evolution of gas content depending on the halo
mass of infall. We summarize our finding as follows:

\begin{itemize}
\item The delay times are in the range of $\tau_\mathrm{d}\sim 0.3-3$~Gyr, while
the fading times are much shorter ($\tau_\mathrm{f}\sim 50-200$~Myr).  This
shows that these simulations are consistent with a delayed-then-rapid
quenching scenario for satellites in group-sized haloes.

\item Halo mass at infall ($M_{\rm h,infall}$) has a strong effect
on the delay time.  Low-$M_{\rm h,infall}$ haloes have large $\tau_\mathrm{d}$,
while high-$M_{\rm h,infall}$ haloes have $\tau_\mathrm{d}\ll 1$~Gyr.
$\tau_\mathrm{d}$ also evolves with redshift, such that they are shorter at $z=1$
than $z=0$.  Finally, galaxies with higher $M_*$ for a given $M_{\rm
h,infall}$ tend to have shorter $\tau_\mathrm{d}$, except for the most massive
$M_{\rm h,infall}>10^{13}M_\odot$ for which the trend is reversed.

\item For the fading time, all values are quite small -- ranging between
50 and 200 Myr -- and typically less than the spacing between individual
snapshots, so they are somewhat uncertain.  There are no clear trends,
except perhaps that $\tau_\mathrm{f}$ seems to be slightly longer
at high $M_{\rm h,infall}$.

\item The delay times for star formation quenching and \HI\ removal
are remarkably similar, and show the same trends as a function of
$M_{\rm h,infall}$ and $M_*$.  This suggests that the \HI\ gas is
being depleted in concert with the star formation being quenched,
broadly indicative of starvation-like mechanisms being the dominant
driver during the delay phase.  Meanwhile, the rapid fading time
in all cases suggests that ram pressure stripping dominates this
final quenching phase, but we lack the time resolution to notice
the expected removal of \HI\ prior to SF quenching.

\item Higher redshift galaxies have higher \HI-richness, which is
independent of the radial distance to the main halo center of mass
at higher redshift, but starts to be dependent on radius at lower
redshift such that closer to the center the galaxies are less
\HI-rich. There is little to no radial dependence of the \HI\ and
H$_2$ contents of galaxies and their efficiency to convert into
stars, and the gas consumption time remains unchanged with respect
to the distance to the main halo at difference cosmic time explored
here ($z\leq2$).

\item Galaxies can more efficiently convert \HI-content into stars
at lower redshift for more massive halo of infall. This scenario
is not present for lower halo mass of infall.  However, a general
trend of higher efficiency  of converting \HI-content into stars in
more massive halo of infall is apparent at all redshift except
perhaps for the highest redshift ($z\sim2$) that displays independent
behaviour with respect to how big the virialised structure the
galaxy falls in.  H$_2$-richness as well as H$_2$-consumption time
are not a function of the halo mass of infall.  The higher
H$_2$-richness and lower H$_2$-consumption time for higher redshift
are barely visible at higher halo mass and even less at lower mass.

\item Comparing to observations, qualitatively we agree with the
trends in \citet{Wetzel-13} and \citet{Fossati-17} for a
delayed-then-rapid quenching with $\tau_\mathrm{d}\gg\tau_\mathrm{f}$.  However, the
actual quenching times (dominated by $\tau_\mathrm{d}$) are shorter by a
factor of two in our simulations than inferred by \citet{Fossati-17}.
However, the work by \citet{Foltz-18} suggests a halo mass dependence of the
quenching timescales with a range of $<1.5$ Gyr which are more
consistent with what our simulations predict, considering they only
looked at cluster member galaxies.

\end{itemize}

Our results demonstrate that, within the group halo mass range
simulated here, the delayed-then-rapid quenching of both star
formation and neutral gas content within satellites is a natural
outcome of the interplay between gas consumption and gas stripping
within realistic haloes.  The actual delay times at face value appear
to be shorter than that inferred from observations, but in future
work we will explore (via mocking up observational procedures) how
the quenching timescales that are inferred from data compare to the
timescales measured directly by tracking the simulated galaxies.
Additionally, a proper statistical comparison would be greatly
improved via a larger sample of zoom simulations, which are
unfortunately quite computationally intensive.  This preliminary
work, nonetheless, demonstrates the power that cosmologically
situated hydrodynamic models can yield valuable insights into the
nature of satellite quenching.

%---------------------------------------------------------------------------------------------------------------------------

\section*{Acknowledgements}
MR and RD acknowledge support from the South African Research Chairs
Initiative and the South African National Research Foundation.
TN acknowledges support from the German Federal Ministry of Education
and Research (BMBF) within the German-South-African collaboration project
01DG 15006. MR acknowledges financial support from Max-Planck-Institut f\"ur
Astrophysik.  Support for MR was also provided by the Square Kilometre
Array post-graduate bursary program.  The zoom~simulations were run
on {\sc draco} which is part of the Max Planck Computing \& Data Facility
(\url{http://www.mpcdf.mpg.de}).

%%%%%%%%%%%%%%%%%%%%%%%%%%%%%%%%%%%%%%%%%%%%%%%%%%

%%%%%%%%%%%%%%%%%%%% REFERENCES %%%%%%%%%%%%%%%%%%

% The best way to enter references is to use BibTeX:

\bibliographystyle{mnras}
\bibliography{paper_bib} % if your bibtex file is called example.bib

\begin{thebibliography}{}
\makeatletter
\relax
\def\mn@urlcharsother{\let\do\@makeother \do\$\do\&\do\#\do\^\do\_\do\%\do\~}
\def\mn@doi{\begingroup\mn@urlcharsother \@ifnextchar [ {\mn@doi@}
  {\mn@doi@[]}}
\def\mn@doi@[#1]#2{\def\@tempa{#1}\ifx\@tempa\@empty \href
  {http://dx.doi.org/#2} {doi:#2}\else \href {http://dx.doi.org/#2} {#1}\fi
  \endgroup}
\def\mn@eprint#1#2{\mn@eprint@#1:#2::\@nil}
\def\mn@eprint@arXiv#1{\href {http://arxiv.org/abs/#1} {{\tt arXiv:#1}}}
\def\mn@eprint@dblp#1{\href {http://dblp.uni-trier.de/rec/bibtex/#1.xml}
  {dblp:#1}}
\def\mn@eprint@#1:#2:#3:#4\@nil{\def\@tempa {#1}\def\@tempb {#2}\def\@tempc
  {#3}\ifx \@tempc \@empty \let \@tempc \@tempb \let \@tempb \@tempa \fi \ifx
  \@tempb \@empty \def\@tempb {arXiv}\fi \@ifundefined
  {mn@eprint@\@tempb}{\@tempb:\@tempc}{\expandafter \expandafter \csname
  mn@eprint@\@tempb\endcsname \expandafter{\@tempc}}}

\bibitem[\protect\citeauthoryear{{Agertz} et~al.,}{{Agertz}
  et~al.}{2007}]{Agertz-07}
{Agertz} O.,  et~al., 2007, \mn@doi [\mnras]
  {10.1111/j.1365-2966.2007.12183.x}, \href
  {http://adsabs.harvard.edu/abs/2007MNRAS.380..963A} {380, 963}

\bibitem[\protect\citeauthoryear{{Bower}, {Benson}, {Malbon}, {Helly}, {Frenk},
  {Baugh}, {Cole}  \& {Lacey}}{{Bower} et~al.}{2006}]{Bower-06}
{Bower} R.~G.,  {Benson} A.~J.,  {Malbon} R.,  {Helly} J.~C.,  {Frenk} C.~S.,
  {Baugh} C.~M.,  {Cole} S.,   {Lacey} C.~G.,  2006, \mn@doi [\mnras]
  {10.1111/j.1365-2966.2006.10519.x}, \href
  {http://adsabs.harvard.edu/abs/2006MNRAS.370..645B} {370, 645}

\bibitem[\protect\citeauthoryear{{Brammer} et~al.,}{{Brammer}
  et~al.}{2012}]{Brammer-12}
{Brammer} G.~B.,  et~al., 2012, \mn@doi [\apjs] {10.1088/0067-0049/200/2/13},
  \href {http://adsabs.harvard.edu/abs/2012ApJS..200...13B} {200, 13}

\bibitem[\protect\citeauthoryear{{Croton} et~al.,}{{Croton}
  et~al.}{2006}]{Croton-06}
{Croton} D.~J.,  et~al., 2006, \mn@doi [\mnras]
  {10.1111/j.1365-2966.2006.09994.x}, \href
  {http://adsabs.harvard.edu/abs/2006MNRAS.367..864C} {367, 864}

\bibitem[\protect\citeauthoryear{{Dav{\'e}}, {Thompson}  \&
  {Hopkins}}{{Dav{\'e}} et~al.}{2016}]{Dave-16}
{Dav{\'e}} R.,  {Thompson} R.,   {Hopkins} P.~F.,  2016, \mn@doi [\mnras]
  {10.1093/mnras/stw1862}, \href
  {http://adsabs.harvard.edu/abs/2016MNRAS.462.3265D} {462, 3265}

\bibitem[\protect\citeauthoryear{{De Lucia}, {Weinmann}, {Poggianti},
  {Arag{\'o}n-Salamanca}  \& {Zaritsky}}{{De Lucia} et~al.}{2012}]{DeLucia-12}
{De Lucia} G.,  {Weinmann} S.,  {Poggianti} B.~M.,  {Arag{\'o}n-Salamanca} A.,
   {Zaritsky} D.,  2012, \mn@doi [\mnras] {10.1111/j.1365-2966.2012.20983.x},
  \href {http://adsabs.harvard.edu/abs/2012MNRAS.423.1277D} {423, 1277}

\bibitem[\protect\citeauthoryear{{Faucher-Gigu{\`e}re}, {Kere{\v s}},
  {Dijkstra}, {Hernquist}  \& {Zaldarriaga}}{{Faucher-Gigu{\`e}re}
  et~al.}{2010}]{Faucher-Giguere-10}
{Faucher-Gigu{\`e}re} C.-A.,  {Kere{\v s}} D.,  {Dijkstra} M.,  {Hernquist} L.,
    {Zaldarriaga} M.,  2010, \mn@doi [\apj] {10.1088/0004-637X/725/1/633},
  \href {http://adsabs.harvard.edu/abs/2010ApJ...725..633F} {725, 633}

\bibitem[\protect\citeauthoryear{{Foltz} et~al.,}{{Foltz}
  et~al.}{2018}]{Foltz-18}
{Foltz} R.,  et~al., 2018, preprint, \href
  {http://adsabs.harvard.edu/abs/2018arXiv180303305F} {} (\mn@eprint {arXiv}
  {1803.03305})

\bibitem[\protect\citeauthoryear{{Fossati} et~al.,}{{Fossati}
  et~al.}{2017}]{Fossati-17}
{Fossati} M.,  et~al., 2017, \mn@doi [\apj] {10.3847/1538-4357/835/2/153},
  \href {http://adsabs.harvard.edu/abs/2017ApJ...835..153F} {835, 153}

\bibitem[\protect\citeauthoryear{{Gabor} \& {Dav{\'e}}}{{Gabor} \&
  {Dav{\'e}}}{2015}]{Gabor-Dave-15}
{Gabor} J.~M.,  {Dav{\'e}} R.,  2015, \mn@doi [\mnras] {10.1093/mnras/stu2399},
  \href {http://adsabs.harvard.edu/abs/2015MNRAS.447..374G} {447, 374}

\bibitem[\protect\citeauthoryear{{Grogin} et~al.,}{{Grogin}
  et~al.}{2011}]{Grogin-11}
{Grogin} N.~A.,  et~al., 2011, \mn@doi [\apjs] {10.1088/0067-0049/197/2/35},
  \href {http://adsabs.harvard.edu/abs/2011ApJS..197...35G} {197, 35}

\bibitem[\protect\citeauthoryear{{Gunn} \& {Gott}}{{Gunn} \&
  {Gott}}{1972}]{Gunn-Gott-72}
{Gunn} J.~E.,  {Gott} III J.~R.,  1972, \mn@doi [\apj] {10.1086/151605}, \href
  {http://adsabs.harvard.edu/abs/1972ApJ...176....1G} {176, 1}

\bibitem[\protect\citeauthoryear{{Hahn} \& {Abel}}{{Hahn} \&
  {Abel}}{2011}]{Hahn-11}
{Hahn} O.,  {Abel} T.,  2011, \mn@doi [\mnras]
  {10.1111/j.1365-2966.2011.18820.x}, \href
  {http://adsabs.harvard.edu/abs/2011MNRAS.415.2101H} {415, 2101}

\bibitem[\protect\citeauthoryear{{Hirschmann}, {De Lucia}, {Wilman},
  {Weinmann}, {Iovino}, {Cucciati}, {Zibetti}  \& {Villalobos}}{{Hirschmann}
  et~al.}{2014}]{Hirschmann-14}
{Hirschmann} M.,  {De Lucia} G.,  {Wilman} D.,  {Weinmann} S.,  {Iovino} A.,
  {Cucciati} O.,  {Zibetti} S.,   {Villalobos} {\'A}.,  2014, \mn@doi [\mnras]
  {10.1093/mnras/stu1609}, \href
  {http://adsabs.harvard.edu/abs/2014MNRAS.444.2938H} {444, 2938}

\bibitem[\protect\citeauthoryear{{Hopkins}}{{Hopkins}}{2015}]{Hopkins-15}
{Hopkins} P.~F.,  2015, \mn@doi [\mnras] {10.1093/mnras/stv195}, \href
  {http://adsabs.harvard.edu/abs/2015MNRAS.450...53H} {450, 53}

\bibitem[\protect\citeauthoryear{{Hopkins}, {Kere{\v s}}, {O{\~n}orbe},
  {Faucher-Gigu{\`e}re}, {Quataert}, {Murray}  \& {Bullock}}{{Hopkins}
  et~al.}{2014}]{Hopkins-14}
{Hopkins} P.~F.,  {Kere{\v s}} D.,  {O{\~n}orbe} J.,  {Faucher-Gigu{\`e}re}
  C.-A.,  {Quataert} E.,  {Murray} N.,   {Bullock} J.~S.,  2014, \mn@doi
  [\mnras] {10.1093/mnras/stu1738}, \href
  {http://adsabs.harvard.edu/abs/2014MNRAS.445..581H} {445, 581}

\bibitem[\protect\citeauthoryear{{Hu}, {Naab}, {Walch}, {Glover}  \&
  {Clark}}{{Hu} et~al.}{2016}]{Hu-16}
{Hu} C.-Y.,  {Naab} T.,  {Walch} S.,  {Glover} S.~C.~O.,   {Clark} P.~C.,
  2016, \mn@doi [\mnras] {10.1093/mnras/stw544}, \href
  {http://adsabs.harvard.edu/abs/2016MNRAS.458.3528H} {458, 3528}

\bibitem[\protect\citeauthoryear{{Kauffmann}, {White}, {Heckman}, {M{\'e}nard},
  {Brinchmann}, {Charlot}, {Tremonti}  \& {Brinkmann}}{{Kauffmann}
  et~al.}{2004}]{Kauffmann-04}
{Kauffmann} G.,  {White} S.~D.~M.,  {Heckman} T.~M.,  {M{\'e}nard} B.,
  {Brinchmann} J.,  {Charlot} S.,  {Tremonti} C.,   {Brinkmann} J.,  2004,
  \mn@doi [\mnras] {10.1111/j.1365-2966.2004.08117.x}, \href
  {http://adsabs.harvard.edu/abs/2004MNRAS.353..713K} {353, 713}

\bibitem[\protect\citeauthoryear{{Kauffmann}, {Li}, {Zhang}  \&
  {Weinmann}}{{Kauffmann} et~al.}{2013}]{Kauffmann-13}
{Kauffmann} G.,  {Li} C.,  {Zhang} W.,   {Weinmann} S.,  2013, \mn@doi [\mnras]
  {10.1093/mnras/stt007}, \href
  {http://adsabs.harvard.edu/abs/2013MNRAS.430.1447K} {430, 1447}

\bibitem[\protect\citeauthoryear{{Kennicutt}}{{Kennicutt}}{1998}]{Kennicutt-98}
{Kennicutt} Jr. R.~C.,  1998, \mn@doi [\apj] {10.1086/305588}, \href
  {http://adsabs.harvard.edu/abs/1998ApJ...498..541K} {498, 541}

\bibitem[\protect\citeauthoryear{{Koekemoer} et~al.,}{{Koekemoer}
  et~al.}{2011}]{Koekemoer-11}
{Koekemoer} A.~M.,  et~al., 2011, \mn@doi [\apjs] {10.1088/0067-0049/197/2/36},
  \href {http://adsabs.harvard.edu/abs/2011ApJS..197...36K} {197, 36}

\bibitem[\protect\citeauthoryear{{Lacerna}, {Contreras}, {Gonz{\'a}lez},
  {Padilla}  \& {Gonzalez-Perez}}{{Lacerna} et~al.}{2018}]{Lancerna-18}
{Lacerna} I.,  {Contreras} S.,  {Gonz{\'a}lez} R.~E.,  {Padilla} N.,
  {Gonzalez-Perez} V.,  2018, \mn@doi [\mnras] {10.1093/mnras/stx3253}, \href
  {http://adsabs.harvard.edu/abs/2018MNRAS.475.1177L} {475, 1177}

\bibitem[\protect\citeauthoryear{{Larson}, {Tinsley}  \& {Caldwell}}{{Larson}
  et~al.}{1980}]{Larson-80}
{Larson} R.~B.,  {Tinsley} B.~M.,   {Caldwell} C.~N.,  1980, \mn@doi [\apj]
  {10.1086/157917}, \href {http://adsabs.harvard.edu/abs/1980ApJ...237..692L}
  {237, 692}

\bibitem[\protect\citeauthoryear{{Lilly} et~al.,}{{Lilly}
  et~al.}{2007}]{Lilly-07}
{Lilly} S.~J.,  et~al., 2007, \mn@doi [\apjs] {10.1086/516589}, \href
  {http://adsabs.harvard.edu/abs/2007ApJS..172...70L} {172, 70}

\bibitem[\protect\citeauthoryear{{McGee}, {Balogh}, {Bower}, {Font}  \&
  {McCarthy}}{{McGee} et~al.}{2009}]{McGee-09}
{McGee} S.~L.,  {Balogh} M.~L.,  {Bower} R.~G.,  {Font} A.~S.,   {McCarthy}
  I.~G.,  2009, \mn@doi [\mnras] {10.1111/j.1365-2966.2009.15507.x}, \href
  {http://adsabs.harvard.edu/abs/2009MNRAS.400..937M} {400, 937}

\bibitem[\protect\citeauthoryear{{Mitra}, {Dav{\'e}}  \& {Finlator}}{{Mitra}
  et~al.}{2015}]{Mitra-15}
{Mitra} S.,  {Dav{\'e}} R.,   {Finlator} K.,  2015, \mn@doi [\mnras]
  {10.1093/mnras/stv1387}, \href
  {http://adsabs.harvard.edu/abs/2015MNRAS.452.1184M} {452, 1184}

\bibitem[\protect\citeauthoryear{{Moster}, {Naab}  \& {White}}{{Moster}
  et~al.}{2018}]{Moster-18}
{Moster} B.~P.,  {Naab} T.,   {White} S.~D.~M.,  2018, \mn@doi [\mnras]
  {10.1093/mnras/sty655}, \href
  {http://adsabs.harvard.edu/abs/2018MNRAS.477.1822M} {477, 1822}

\bibitem[\protect\citeauthoryear{{Muratov}, {Keres}, {Faucher-Giguere},
  {Hopkins}, {Quataert}  \& {Murray}}{{Muratov} et~al.}{2015}]{Muratov-15}
{Muratov} A.~L.,  {Keres} D.,  {Faucher-Giguere} C.-A.,  {Hopkins} P.~F.,
  {Quataert} E.,   {Murray} N.,  2015, ArXiv e-prints:1501.03155, \href
  {http://adsabs.harvard.edu/abs/2015arXiv150103155M} {}

\bibitem[\protect\citeauthoryear{{Naab} \& {Ostriker}}{{Naab} \&
  {Ostriker}}{2017}]{Naab-17}
{Naab} T.,  {Ostriker} J.~P.,  2017, \mn@doi [\araa]
  {10.1146/annurev-astro-081913-040019}, \href
  {http://adsabs.harvard.edu/abs/2017ARA%26A..55...59N} {55, 59}

\bibitem[\protect\citeauthoryear{{Oman} \& {Hudson}}{{Oman} \&
  {Hudson}}{2016}]{Oman-Hudson-16}
{Oman} K.~A.,  {Hudson} M.~J.,  2016, \mn@doi [\mnras] {10.1093/mnras/stw2195},
  \href {http://adsabs.harvard.edu/abs/2016MNRAS.463.3083O} {463, 3083}

\bibitem[\protect\citeauthoryear{{Oser}, {Ostriker}, {Naab}, {Johansson}  \&
  {Burkert}}{{Oser} et~al.}{2010}]{Oser-10}
{Oser} L.,  {Ostriker} J.~P.,  {Naab} T.,  {Johansson} P.~H.,   {Burkert} A.,
  2010, \mn@doi [\apj] {10.1088/0004-637X/725/2/2312}, \href
  {http://adsabs.harvard.edu/abs/2010ApJ...725.2312O} {725, 2312}

\bibitem[\protect\citeauthoryear{{Peng} et~al.,}{{Peng} et~al.}{2010}]{Peng-10}
{Peng} Y.-j.,  et~al., 2010, \mn@doi [\apj] {10.1088/0004-637X/721/1/193},
  \href {http://adsabs.harvard.edu/abs/2010ApJ...721..193P} {721, 193}

\bibitem[\protect\citeauthoryear{{Planck} et~al.,}{{Planck} et~al.}{2016}]{-16}
{Planck} et~al., 2016, \mn@doi [\aap] {10.1051/0004-6361/201525830}, \href
  {http://adsabs.harvard.edu/abs/2016A%26A...594A..13P} {594, A13}

\bibitem[\protect\citeauthoryear{{Rafieferantsoa} \&
  {Dav{\'e}}}{{Rafieferantsoa} \& {Dav{\'e}}}{2018}]{Rafieferantsoa-18}
{Rafieferantsoa} M.,  {Dav{\'e}} R.,  2018, \mn@doi [\mnras]
  {10.1093/mnras/stx3293}, \href
  {http://adsabs.harvard.edu/abs/2018MNRAS.475..955R} {475, 955}

\bibitem[\protect\citeauthoryear{{Rafieferantsoa}, {Dav{\'e}},
  {Angl{\'e}s-Alc{\'a}zar}, {Katz}, {Kollmeier}  \&
  {Oppenheimer}}{{Rafieferantsoa} et~al.}{2015}]{Rafieferantsoa-15}
{Rafieferantsoa} M.,  {Dav{\'e}} R.,  {Angl{\'e}s-Alc{\'a}zar} D.,  {Katz} N.,
  {Kollmeier} J.~A.,   {Oppenheimer} B.~D.,  2015, \mn@doi [\mnras]
  {10.1093/mnras/stv1933}, \href
  {http://adsabs.harvard.edu/abs/2015MNRAS.453.3980R} {453, 3980}

\bibitem[\protect\citeauthoryear{{Rahmati}, {Pawlik}, {Rai\v{c}evi\`{c}}  \&
  {Schaye}}{{Rahmati} et~al.}{2013}]{Rahmati-13}
{Rahmati} A.,  {Pawlik} A.~H.,  {Rai\v{c}evi\`{c}} M.,   {Schaye} J.,  2013,
  \mn@doi [\mnras] {10.1093/mnras/stt066}, \href
  {http://adsabs.harvard.edu/abs/2013MNRAS.430.2427R} {430, 2427}

\bibitem[\protect\citeauthoryear{{Schmidt}}{{Schmidt}}{1959}]{Schmidt-59}
{Schmidt} M.,  1959, \mn@doi [\apj] {10.1086/146614}, \href
  {http://adsabs.harvard.edu/abs/1959ApJ...129..243S} {129, 243}

\bibitem[\protect\citeauthoryear{{Simha}, {Weinberg}, {Dav{\'e}}, {Gnedin},
  {Katz}  \& {Kere{\v s}}}{{Simha} et~al.}{2009}]{Simha-09}
{Simha} V.,  {Weinberg} D.~H.,  {Dav{\'e}} R.,  {Gnedin} O.~Y.,  {Katz} N.,
  {Kere{\v s}} D.,  2009, \mn@doi [\mnras] {10.1111/j.1365-2966.2009.15341.x},
  \href {http://adsabs.harvard.edu/abs/2009MNRAS.399..650S} {399, 650}

\bibitem[\protect\citeauthoryear{{Somerville} \& {Dav{\'e}}}{{Somerville} \&
  {Dav{\'e}}}{2015}]{Somerville-15r}
{Somerville} R.~S.,  {Dav{\'e}} R.,  2015, \mn@doi [\araa]
  {10.1146/annurev-astro-082812-140951}, \href
  {http://adsabs.harvard.edu/abs/2015ARA%26A..53...51S} {53, 51}

\bibitem[\protect\citeauthoryear{{Voit}}{{Voit}}{2005}]{Voit-05}
{Voit} G.~M.,  2005, \mn@doi [Advances in Space Research]
  {10.1016/j.asr.2005.02.042}, \href
  {http://adsabs.harvard.edu/abs/2005AdSpR..36..701V} {36, 701}

\bibitem[\protect\citeauthoryear{{Weinmann}, {van den Bosch}, {Yang}  \&
  {Mo}}{{Weinmann} et~al.}{2006}]{Weinmann-06}
{Weinmann} S.~M.,  {van den Bosch} F.~C.,  {Yang} X.,   {Mo} H.~J.,  2006,
  ArXiv Astrophysics e-prints, \href
  {http://adsabs.harvard.edu/abs/2006astro.ph..7585W} {}

\bibitem[\protect\citeauthoryear{{Wetzel}, {Tinker}, {Conroy}  \& {van den
  Bosch}}{{Wetzel} et~al.}{2013}]{Wetzel-13}
{Wetzel} A.~R.,  {Tinker} J.~L.,  {Conroy} C.,   {van den Bosch} F.~C.,  2013,
  \mn@doi [\mnras] {10.1093/mnras/stt469}, \href
  {http://adsabs.harvard.edu/abs/2013MNRAS.432..336W} {432, 336}

\bibitem[\protect\citeauthoryear{{Wetzel}, {Tollerud}  \& {Weisz}}{{Wetzel}
  et~al.}{2015}]{Wetzel-15}
{Wetzel} A.~R.,  {Tollerud} E.~J.,   {Weisz} D.~R.,  2015, \mn@doi [\apjl]
  {10.1088/2041-8205/808/1/L27}, \href
  {http://adsabs.harvard.edu/abs/2015ApJ...808L..27W} {808, L27}

\bibitem[\protect\citeauthoryear{{Wisnioski} et~al.,}{{Wisnioski}
  et~al.}{2015}]{Wisnioski-15}
{Wisnioski} E.,  et~al., 2015, \mn@doi [\apj] {10.1088/0004-637X/799/2/209},
  \href {http://adsabs.harvard.edu/abs/2015ApJ...799..209W} {799, 209}

\bibitem[\protect\citeauthoryear{{van Gorkom}}{{van
  Gorkom}}{2004}]{vanGorkom-04}
{van Gorkom} J.~H.,  2004, Clusters of Galaxies: Probes of Cosmological
  Structure and Galaxy Evolution, \href
  {http://adsabs.harvard.edu/abs/2004cgpc.symp..305V} {p.~305}

\bibitem[\protect\citeauthoryear{{van Gorkom} et~al.,}{{van Gorkom}
  et~al.}{2003}]{vanGorkom-03}
{van Gorkom} J.~H.,  et~al., 2003, \mn@doi [\apss] {10.1023/A:1024607103456},
  \href {http://adsabs.harvard.edu/abs/2003Ap%26SS.285..219V} {285, 219}

\makeatother
\end{thebibliography}

%%%%%%%%%%%%%%%%%%%%%%%%%%%%%%%%%%%%%%%%%%%%%%%%%%

% Don't change these lines
%\bsp	% typesetting comment
\label{lastpage}
\end{document}